\documentclass[12pt]{article}

\begin{document}

\def\beq{\begin{equation}}
\def\eeq{\end{equation}}
\def\beqa{\begin{eqnarray}}
\def\eeqa{\end{eqnarray}}
\def\eq#1{Eq.~(\ref{#1})}
\newcommand{\secn}[1]{Section~\ref{#1}}

\begin{titlepage}
\rightline{DFTT 35/02}
\rightline{\hfill October 2002}

\vskip 2cm

\centerline{\Large \bf On power corrections to event shapes\footnote{Invited
talk at the XIV Italian Meeting on High Energy Physics, Parma,
April 2002.}}
 
\vskip 2cm

\centerline{\bf Lorenzo Magnea\footnote{e-mail: 
{\tt magnea@to.infn.it}}}
\centerline{\sl Dipartimento di Fisica Teorica, Universit\`a di Torino}
\centerline{\sl and INFN, Sezione di Torino,}
\centerline{\sl Via P.~Giuria 1, I--10125 Torino, Italy}
 
\vskip 2cm
 
\begin{abstract}

\noindent Recent work on the theme of power corrections in
perturbative QCD is briefly reviewed, with an emphasis on event shapes
in $e^+ e^-$ annihilation. The factorization of soft gluon effects is
the main tool: it leads to resummation, and thus highlights the
limitations of perturbation theory, pointing to nonperturbative
corrections whose size can be estimated. Power corrections can be
resummed into shape functions, for which QCD--based models are
available.  Theoretical progress is closing in on the nonperturbative
frontier.

\end{abstract}

\end{titlepage}

\newpage

\section{Introduction}
\label{intro}

It has been known since the early days of QCD~\cite{thooft} that the
perturbative expansion for IR safe observables is at best asymptotic,
and in fact not even Borel summable.  This is good news: it would be
very surprising if perturbation theory alone could give a
well--defined answer in a theory such as QCD, where nonperturbative
phenomena govern the spectrum of physical states; an ambiguity in the
perturbative answer tells us that confinement physics must, at some
level, be present and relevant. Under the mild assumption that QCD is
a consistent theory, we can do more: gauging the size of the
uncertainty in the perturbative answer, we can estimate the impact of
the dominant nonperturbative effects.

The basic tools for this analysis are the factorization and the
resummation of soft gluon effects. Soft gluon emission has a universal
character, and factorizes from the hard part of scattering amplitudes
and cross sections; factorization in turn implies that the dominant
(logarithmic) contribution of soft gluons can be computed to all
orders. These computations display explicitly the asymptotic nature
of the perturbative series and can be used to study power--suppressed
corrections.

Studies of power corrections with perturbative methods have generated
a vast literature in the past several years~\cite{martin}, and have
reached an impressive degree of phenomenological success. In this
short review I will focus mostly on theoretical results obtained for
event shapes in $e^+ e^-$ annihilation, trying to put them in the
perspective of the most recent developments.

\section{\hspace{-2mm} Perturbative windows on power corrections}
\label{perwi}

QCD resummations typically yield expressions of the general form
\beq
f_a (q^2) = \int_0^{q^2} \frac{d k^2}{k^2} (k^2)^a \alpha_s(k^2)~.
\label{genpow1} 
\eeq
Such expressions are ill--defined in perturbative QCD, because of the
Landau pole in the running coupling at $k^2 = \Lambda^2$. Expanding in
powers of $\alpha_s (q^2)$, one finds that the coefficients of the
expansion grow factorially. The size of the ambiguity in the result
can be gauged by taking the residue of the pole: one finds that it is
suppressed by a power of the hard scale, $\delta f_a/f_a \propto
\left(\Lambda^2/q^2 \right)^a$.

A classic example~\cite{martin} is the resummation of fermion
bubble insertions into a gluon line. Let $\Pi (k^2)$ denote the
fermionic part of the gluon vacuum polarization, and let $\sigma (x)$
be a generic (possibly weighted) cross section depending on some
kinematic variable $x$.  Summing over insertions of $\Pi$ in the
single gluon contribution to $\sigma$ gives an expression of the form
\beq
\sigma (x) = \int \frac{d k^2}{k^2} \frac{1}{|1 + \Pi(k^2)|^2} 
\widehat{\sigma} \left(x, \frac{k^2}{q^2} \right)~,
\label{sigmal}
\eeq
where $\widehat{\sigma}$ is the virtuality distribution of the emitted
gluon.  If the $n_f$ dependence of the answer is reinterpreted as the
abelian contribution to the running of the coupling, \eq{sigmal}
displays the Landau pole in the integration over gluon virtuality.

The resummation of multiple soft gluon emission near the boundary of 
phase space leads to similar conclusions. Consider, for example,
the Laplace transform of the thrust distribution~\cite{CTTW}. It can be 
shown to exponentiate in the form
\beq
\int_0^1 dt\, {\rm e}^{-\nu t} \frac{d\sigma}{dt} = 
{\rm e}^{-S(\nu,Q)}\,,
\label{Laplace}
\eeq
where the function $S$ contains all the singular logarithmic dependence 
on $t = 1 - T$, and can be written (to NLL accuracy) as
\beq
S(\nu,Q) = \int_0^1 {d\alpha \over \alpha}\; \left( 1 - 
{\rm e}^{-\nu\alpha} \right) \left [
\int_{\alpha^2 q^2}^{\alpha q^2} {dk_\perp^2 \over\ k_\perp^2}\;
\Gamma \left(\alpha_s(k_\perp^2)\right) + B \left( \alpha_s 
(\alpha q^2) \right) \right] 
\label{lapexp}
\eeq
In this case the Landau pole obstructs the integration over transverse
momentum.

A variety of regularizations for the Landau singularity have been
proposed, (IR cutoff~\cite{coso}, principal value
prescription~\cite{princi}, regular IR continuation of the running
coupling~\cite{DMW}, dimensional regularization~\cite{medr}, choice of
contour in the inverse Laplace transform~\cite{CMNT}). As far as the
parametric size of the power corrections ({\it i.e.} the power of $q$
with which they scale) is concerned, all these prescriptions
(sometimes after some debate) lead to the same conclusion. They may
differ when more detailed prediction are attempted.

A benchmark to verify the reliability of the methods based on
different resummations is the application to observables for which
nonperturbative information is available through other means,
typically because of the applicability of the OPE. Along these lines
resummation and OPE methods have been shown to give compatible results
for the size of the correction in all tested cases (ranging from the
beautiful results of David~\cite{dav} on the nonlinear $\sigma$ model
in $d = 2$, to the classic analysis of the total annihilation cross
section by Mueller~\cite{muel}, to the more recent results on deep
inelastic structure functions~\cite{koga}). Making more detailed
predictions, going beyond the identification of the leading power
correction, it is necessary to rely upon the assumption of {\it
ultraviolet dominance}~\cite{BBM}. To introduce the idea, consider the
OPE for an inclusive cross section, say a DIS structure function in
Mellin space
\beqa
F_a (N, q^2) & = & C^i_{2,a} (N, \mu_f, q^2) \langle O^{(2)}_i (N, \mu_f) 
\rangle \nonumber \\ & + & \frac{1}{q^2} C^i_{4,a} (N, \mu_f, q^2) 
\langle O^{(4)}_i (N, \mu_f) \rangle + \ldots~.
\label{disope} 
\eeqa
Here the factorization scale $\mu_f$ is an IR cutoff for the
coefficient functions $C_i$, but it is an ultraviolet cutoff for the
operator matrix elements. Physical quantities must not depend on
$\mu_f$, and in fact the cancellation of the {\it logarithmic}
dependence on $\mu_f$, {\it within} a given twist, is enforced by
Altarelli--Parisi equations.  One observes, however, that the
coefficient functions $C_i$ have an ambiguous power dependence on
$\mu_f$, of IR origin, due to the divergence of their perturbative
expansion. At the same time, higher twist operator matrix elements
have power--like UV divergences, so that they mix under
renormalization with lower twist operators. These two facts make the
twist separation ambiguous, and the ambiguity can only be disentangled
once the same regularization prescription is chosen for both $C_i$ and
$\langle O_i \rangle$. What renormalon models and resummations are
trying to do is to predict the exact dependence of higher twist
operator matrix elements on kinematical variables (such as $N$ above),
based on their renormalization properties, {\it i.e.} on their UV
behavior. These prediction will be correct to the extent that the
matrix elements are ``UV dominated''. Evidence supporting the idea of
UV dominance has recently been presented~\cite{koga} in the case of 
DIS structure functions, in the elastic limit $x \to 1$. In that limit 
matrix elements of operators of twist $4$ and $6$ are dominated by 
parton configurations mimicking twist $2$, so one can explicitly verify 
that the $N$ dependence of the perturbative ambiguity at 
twist $2$ (and large $N$) is in fact cancelled by the corresponding 
ambiguity of the higher--twist operator matrix elements.

To go beyond the OPE, it is necessary to either rely upon renormalon
models, or to take the viewpoint suggested by the factorization
properties underlying soft gluon resummations. In the following, I
will use factorization to define a way to parametrize power
corrections to event shapes (the shape function),
following~\cite{koste}; starting from this general parametrization it
is possible to recover the results of renormalon calculus, which in
turn can be viewed as a QCD--motivated model for the shape function.

\section{Two--jet limit and shape functions}
\label{shafu}

Generic IR safe observables in production processes cannot be
described with the OPE: they are {\it weighted cross
sections}. Specifically, consider the distribution of an event shape
$e$, chosen so that the two--jet limit correspond to $e \to 0$ (for
example, $e = 1 - T, C, \rho_J, \ldots$).  If the explicit expression
for $e$ in terms of the momenta of final state particles is $e = E_m
(p_1, \ldots, p_m)$, one defines
\beq
\frac{d \sigma}{d e} = \frac{1}{2 q^2} \sum_m \int d {\rm LIPS}_m~
\overline{\left| {\cal M}_m \right|^2} ~\delta \left(e - E_m (p_1, 
\ldots, p_m) \right)~,
\label{diste}
\eeq
as well as the `radiation function', $R(e) \equiv \int_0^e d e' d
\sigma/d e'$.

Soft gluon effects dominate the cross section in the two--jet region.
Final state gluons are forced to be either soft, or collinear to the
two back--to--back jets; as a consequence, the distribution factorizes
under a Laplace (or Mellin) transform, schematically as $\sigma(e,
q^2) = J_1(q^2 e)*J_2 (q^2 e)*S(q^2 e^2)*H(q^2)$. There are, in fact,
two relevant mass scales in the $e \to 0$ limit: the squared invariant
mass of the jets, which vanishes as $q^2 e$, and the (squared) total
energy carried by soft gluons, which vanishes as $q^2 e^2$. The
factorization is valid, within perturbation theory and for $e \to 0$,
in the range $q^2 >> q^2 e >> q^2 e^2 >> \Lambda^2$.  If we assume
that we can rely upon the same factorization in the extended range
$q^2 >> q^2 e >> q^2 e^2 \sim \Lambda^2$, we can use the perturbative
results to parametrize power corrections of the type $(\Lambda/(q
e))^p$. 

Physically, the picture emerging from factorization is equivalent to
the phenomenological `tube model'~\cite{webbe}: neglecting inverse
powers of $q^2 e$, one observes that jet masses are insensitive to the
transverse components of soft gluon momenta, and are linearly shifted
by an amount proportional to the total light--cone momentum carried by
soft gluons,
\beq
\left. M^2_i \right|_{NP} = \left. M^2_i \right|_{PT} + \epsilon_i q~,
\label{jema}
\eeq
where $i$ labels the two jets and $\epsilon_i$ is the sum of soft
gluon momentum components along the $i$-th jet direction. 
At least for event shapes vanishing with $M^2_i$, the effect of 
soft gluons on the perturbative distribution will take the form of a 
convolution in the light--cone components $\epsilon_i$. Introducing an
IR cutoff $\mu$, one writes
\beq
R (e) = \left. \int_0^\mu d \epsilon_1 d\epsilon_2 f(\epsilon_1,\epsilon_2)
R_{PT} (e) ~\right|_{M^2_i \rightarrow M^2_i + \epsilon_i q}~,
\label{RHp}
\eeq
where $f(\epsilon_1, \epsilon_2)$ is the most general form of the
announced shape function. Note that it does not depend on the hard
scale $q$.  For shape variables such as $C$ or $t$, which depend only
on the sum of two jet masses in the two--jet limit, one can simplify
the convolution by introducing
\beq
f (\epsilon) = \int_0^\mu d \epsilon_1 d\epsilon_2 f(\epsilon_1,\epsilon_2)
\delta(\epsilon - \epsilon_1 - \epsilon_2)~.
\label{sincon}
\eeq
Factorization further provides an explicit, if formal,
operator expression for the shape function $f$. In the limit we are
considering, in fact, soft gluon emission is well approximated by
replacing the two back--to--back jets by eikonal lines. One can then
treat the joint distribution of $\epsilon_{1,2}$ as a generic event
shape, according to \eq{diste}, but replacing the matrix element 
${\cal M}_m$ with its eikonal counterpart. Essentially
\beq
f (\epsilon_1, \epsilon_2) = \frac{1}{2 q^2} \sum_m \int d 
{\rm LIPS}^{\rm eik}_m~ \overline{\left| {\cal M}^{\rm eik}_m \right|^2} 
\delta \left(\epsilon_1 - k^{\rm soft}_1 \right)
\delta \left(\epsilon_2 - k^{\rm soft}_2 \right)~,
\label{opdef}
\eeq
where $d {\rm LIPS}^{\rm eik}_m$ is the $m$--particle phase space with
fixed eikonal lines, and $k^{\rm soft}_i$ is the total soft gluon
light--cone momentum flowing into hemisphere $i$.
Such eikonal cross sections are matrix elements of Wilson lines, so the
shape function can ultimately be expressed in terms of correlators of 
the energy--momentum tensor at large distances, in the presence of eikonal 
sources.

To compare the results of this approach with renormalon calculations,
it is useful to see how the shape function emerges from a resummed
perturbative calculation. Consider the expression for thrust,
\eq{lapexp}.  There, small $t$ is associated with large $\nu$,
and the dominant contributions to the integral arise from the region
$\alpha \sim 1/\nu \sim t$. The leading power corrections that we are
trying to resum are then related to the IR behavior of the coupling
through the cusp anomalous dimension $\Gamma$, and one can isolate 
them by interchanging the order of integration, and introducing a 
transverse momentum cutoff. One defines
\begin{eqnarray}
S_{NP}(\nu/q,\mu) & = & \int_0^{\mu^2} \frac{d k_\perp^2}{k_\perp^2} 
\Gamma \left( \alpha_s (k_\perp^2) \right) \int_{k_\perp^2/q^2}^{k_\perp/q}
\frac{d \alpha}{\alpha} \left(1 - {\rm e}^{- \nu \alpha} \right)
\nonumber \\ & = & 
\sum_{n = 1}^\infty \frac{1}{n!} \left( \frac{\nu}{q}
\right)^n \lambda_n(\mu^2)~,
\label{sudpar}
\end{eqnarray}
where terms suppressed by powers of $\nu/q^2$ have been neglected and
\beq
\lambda_n(\mu^2) = \frac{1}{n} \int_0^{\mu^2} d k_\perp^2 k_\perp^{n-2} 
\Gamma \left( \alpha_s (k_\perp^2) \right)
\label{lampar}
\eeq
are an infinite set of dimensionful nonperturbative 
parameters, controlling power corrections of decreasing size, but always 
of the desired form $1/(t q)^n$. They can be organized into a shape 
function via a Laplace transform
\beq
\exp \left( - S_{NP}(\nu/q,\mu) \right) \equiv
\int_0^\infty d \epsilon {\rm e}^{- \nu \epsilon/q} f_t(\epsilon, 
\mu)~. 
\label{shaf1}
\eeq
Within this perturbative framework, it is easy to make contact with
the dispersive approach~\cite{DMW}, whose results are recovered at
leading power. Specifically,
\begin{itemize}
\item the dispersive approach predicts {\it moments} of event shapes
in terms of a single nonperturbative parameter $\alpha_0$. This
parameter is equivalent (in fact, in a suitable factorization scheme,
proportional) to $\lambda_1$. From the point of view of the shape
function, one can show that the parameters $\lambda_n$ with $n >1$
contribute to power corrections to moments only at the level of
subleading corrections, $1/q^p$ with $p>1$. Power corrections for the
moments of the distribution are then, in fact, well approximated by
retaining only $\lambda_1$: it is only in the deep nonperturbative
regime $e \sim \Lambda/q$ that subleading power corrections play a
significant role. It is interesting to note~\cite{koste2} that within
the dispersive approach central moments (such as $\Delta^2 t = \langle
t^2\rangle - \langle t \rangle^2$) receive no leading power
corrections, whereas these are nonvanishing in shape function fits.
\item In the dispersive approach one needs to include a correction
accounting for the {\it correlations} between emissions in
different hemispheres~\cite{nase,BBM,mila,DMS}. The shape function
accounts for this effect by a lack of factorizability: if one writes
$f(\epsilon_1, \epsilon_2) = g(\epsilon_1) g(\epsilon_2) + \delta
f(\epsilon_1, \epsilon_2)$, the $\delta f$ term is responsible for
correlations. Clearly, renormalon calculations provide a model
estimate of this effect, whereas from the point of view of the shape
function this will emerge as a result of fits to data.
\item Tube and renormalon models predict that the leading power
correction shifts event shape {\it distributions} by an amount
proportional to the energy carried by soft gluons. This prediction is
recovered by the shape function, which introduces additional smearing
due to subleading power corrections, as easily seen by using
eq.~\ref{shaf1}, and as described in detail in Ref.~\cite{koste}.
\end{itemize}

\section{Dressed gluon exponentiation}
\label{dregl}

It is clearly desirable to merge the information that can be extracted
from soft gluon resummation (which expresses the dynamical effects of
perturbative multiple soft gluon emission) with the results of the
dispersive approach (which models the effects of nonperturbative
single gluon emission).  This merging is achieved~\cite{Gardi} by the
method of dressed gluon exponentiation (DGE).

To summarize the main features of the method, pick as a shape 
variable a jet mass $\rho$. Schematically, one proceeds as 
follows~\cite{gara}.

\begin{itemize}
\item The first step is the computation of the characteristic function
of the dispersive method (that is, the differential cross section for
the emission of a gluon with virtuality $k^2$) for the observable at hand,
in the Sudakov limit. In this limit, one retains only terms that are
singular as $\rho \to 0$ when $k^2/(\rho q^2)$ is kept finite, as
these are the only terms contributing to the Sudakov logarithms in the
two--jet limit.  For the jet mass one finds, setting $\xi \equiv
k^2/q^2$,
\beq
\left. \dot{\cal F}(\rho, \xi) \right\vert_{\log} =
\frac{2}{\rho}-\frac{\xi}{\rho^2}-\frac{\xi^2}{\rho^3}~.
\label{hjm1}
\eeq
\item Within the dispersive approach, the characteristic function must
be integrated over virtuality with a weight given by the running
coupling.  Here one encounters the Landau pole. One possibility to
define the integration is to turn to a Borel representation of the
coupling~\cite{begru}.  In the present case, to achieve NLL accuracy
it is crucial to define the coupling so that it satisfies Altarelli--
Parisi evolution at NLO, at least for the singular contributions near
threshold. This is the ``gluon bremsstrahlung''
coupling~\cite{arica}. Performing the integration over gluon
virtuality, the result for the distribution is of the form
\beq
\frac{d\sigma}{d \rho} =
\frac{C_F}{2\beta_0}\int_0^{\infty} du B (u,\rho)
\exp\left(-{{u} \ln (q^2/\bar{\Lambda}^2}) \right) 
\frac{\sin \pi u }{\pi u} \bar{A}_B(u)~.
\label{Borel_rep}
\eeq
Here $\bar{A}_B(u)$ is the Borel transform of the chosen coupling, the
factor $\sin \pi u/(\pi u)$ arises from having taken the timelike
discontinuity of the coupling, as prescribed by the dispersive method,
and $B (u,\rho)$ is the Borel function obtained by integrating
\eq{hjm1} over virtuality with the appropriate weight.
\item The key step to tie this single gluon result to resummation is to
use the dressed gluon distribution as kernel of exponentiation. One writes
the Laplace transform of the jet mass distribution as
\beq
\ln J(\nu,q)=\int_0^1 \frac{d\sigma}{d\rho}
\left(e^{-\nu \rho}-1\right)d\rho~, 
\label{ln_J_nu} 
\eeq
and substitutes in this expression the single dressed gluon result, 
given by \eq{Borel_rep}.
\item At this point, the Borel representation of the exponent suggests
a pattern of exponentiated power corrections. The integral transform
associated with the exponentiation turns the simple pole structure of
the one--loop Borel function $B (u,\rho)$ into a much richer one for
the transformed function $\tilde{B}(u, \nu)$. One can use this
structure to model the pattern of power corrections. The radiation
function can be reconstructed by inverting the Laplace transform,
using
\beq
R (\rho) = \int_C \frac{d\nu}{2\pi i\nu} 
\exp \left[\nu \rho + \ln J^{\rm PT}(\nu,q) + \ln J^{\rm NP}(\nu,q)  
\right]~,
\label{invlap}
\eeq
with a nonperturbative contribution
\beq
\ln J^{\rm NP}(\nu, q) = - \sum_{n=1}^{\infty}
\widehat{\lambda}_n \frac{1}{n!} \left(\frac{\nu \Lambda}{q}\right)^n,
\label{powser}
\eeq
In this approach, $\widehat{\lambda}_n$ are still free parameters,
however the structure of the Borel transform may fix relations between
them, or other general properties. For example, in the case of the jet
mass $\rho$ one is led to conclude that $\widehat{\lambda}_{2 k} = 0$.
\item These results can again be interpreted in terms of a shape function,
writing $J^{\rm NP} (\nu, q)$ as in \eq{shaf1}.
Now however the structure of the moments is derived from the Borel 
representation. 
\end{itemize}
Dressed gluon exponentiation provides a useful tool to merge the
predictions of soft gluon resummation with a parametrization of power
corrections dictated by renormalons. The results of resummation are
reproduced to NLL accuracy, but further predictions are made: in fact,
the size of all subleading perturbative logarithms is predicted in the
large $n_f$ limit, and found to grow factorially. This fact, perhaps
not unexpected, in view of the renormalon singularities of the original
resummed expression, sets limits on the validity of resummations
performed in terms of successive towers of logarithms, and provides an
independent tool to estimate the range in which they are applicable.
As far as power corrections are concerned, DGE shows that in the
two--jet limit at least a class of them exponentiates together with
Sudakov logarithms. Further, DGE provides a definite regularization
prescription to handle resummed perturbative expressions at power
accuracy, which may be useful to check the assumption of ultraviolet
dominance whenever the explicit operator form of power suppressed
contributions is known, as was done for DIS in Ref.~\cite{koga}.
Finally, as announced, renormalons make specific predictions on the
structure of the shape function for a given observable, which in
principle can be tested against data.

\section{Phenomenology}
\label{pheno}

The shape function approach and DGE were applied to selected event
shapes in~\cite{koste2} and~\cite{gara}. For lack of space, I must
refer the reader to the original papers for details, however some
observations are in order.

First of all, as pointed out above, DGE puts constraints on the form
of the shape function, which may depend on how renormalon calculus is
implemented. For example, Korchemsky and Tafat, in Ref.~\cite{koste2},
include in their fit the effect of correlations between hemispheres by
picking a gaussian ansatz for the shape function with a
non--factorizable term; this term is found to be important for the
fit. Gardi and Rathsman~\cite{gara}, on the other hand, do not include
in their analysis the two--loop correction responsible for
correlations within the dispersive approach, and argue that its
contribution should be negligible. Such differences are perhaps
difficult to settle when working with multiparameter fits, however it
seems that a more extensive comparison with data for different event
shapes should settle this issue.

A second important point is the fact that if the strong coupling
$\alpha_s (M_z)$ is also included as a fit parameter, the results
point to a value which is rather low with respect to the world average
(typically $\alpha_s (M_z) \sim 0.110$). This should perhaps be
considered as a cause of moderate concern: it could mean that the
current expressions for the shape function are not fully adequate to
describe the effect of power corrections in the peak region (and thus
they pull the fit towards lower values of $\alpha_s$), or it could imply
that current assumptions about the theoretical error to be associated
with $\alpha_s (M_z)$ underestimate the effect of higher order and
power corrections.

A final observation of considerable relevance for phenomenology was
made by Salam and Wicke~\cite{sawi}, and it concerns the relationship
between the theoretical calculations described so far and experimental
data. Briefly stated, the point is that while all models of power
corrections are derived in massless QCD, event shapes are measured
using the massive particles produced by hadronization.  Salam and
Wicke observed that the difference between massless and massive
definitions of event shapes induces non--universal power corrections
of the same parametric size ($\Lambda/q$) as conventional ones.  To
disentangle these mass effects from the universal features of soft
gluon emission, for example when fitting nonperturbative parameters,
it is necessary to specify a scheme to connect massless QCD
computations and measured event shapes. The procedure followed this
far, which ignores mass effects in the definition of events shapes,
also constitutes a possible scheme; in this scheme, however, different
event shapes are treated differently, and non--universal,
mass--related power corrections should be fitted separately for
different observables.  Other schemes have been proposed: notably, the
``$E$--scheme'' uses measured energies $E_i$ in the definition of the
event shape, and rescales three--momenta by $E_i/p_i$; in this scheme,
three--momentum is not conserved, however it can be shown that
non--universal power corrections vanish in a tube model
calculation. Another interesting possibility is the ``decay scheme'',
in which all measured particles are forced to decay into massless
particles via a Monte Carlo interface; not all decays are strong, nor
realistic, however the method is interesting because it begins to
address in some detail the issue of possible double countings, which
arises when parton--level, QCD--based predictions for power
corrections are used in conjunction with other hadronization models
such as those implemented in Monte Carlo evolution codes.

Mass--related power corrections of the type described in Ref.~\cite{sawi} 
are further logarithmically enhanced by effects related to hadron 
multiplicity in the final state. Model calculations using the hypothesis 
of local parton--hadron duality suggest
\beq
\delta_m \langle e \rangle = c_e \frac{\Lambda}{q} 
\left(\log \frac{q}{\Lambda} \right)^A~; \quad A = \frac{4 N_c}{b_0} 
\sim 1.6~,
\label{sawi}
\eeq
where $\delta_m \langle e \rangle$ is the non--universal power
correction to the average of event shape $e$, and $c_e$ a fitted
coefficient. It should be noted that conventional power corrections
will also generically be enhanced by logarithms, however not much
attention has so far been devoted to study their effects.

\section{Perspectives}
\label{persp}

After more than two decades of studies, event shape distributions
remain at the forefront of theoretical and experimental QCD analyses.
In a single graph, a typical distribution takes us from a completely
perturbative regime ($e \sim 1$), where, at least at high energy,
fixed order calculations apply, to a region dominated by
nonperturbative effects ($e \sim \Lambda/q$), where all power
corrections become important.  Successive improvements in our
theoretical tools are leading to a QCD--based understanding of the
entire distribution, including the peak region dominated by two--jet
configurations.

Resummed QCD amplitudes point beyond perturbation theory, and the
effects of nonpertubative corrections can be conveniently parametrized
by a new class of nonperturbative functions, the shape functions,
which provide a general framework for studies of power
corrections. Like parton distributions, they must be fitted from data,
however different QCD--motivated models suggest somewhat different
functional forms~\cite{gara,BKS}. The results of the dispersive
approach are recovered, and the universality of the leading power
corrections placed in a wider perspective. Renormalon calculus can be
merged with Sudakov resummation via dressed gluon exponentiation,
providing a model for the shape function. Hadronization effects
generate mass--related, log--enhanced power corrections which must be
separately understood: a defining scheme for event shapes must be
chosen before a detailed phenomenological analysis is attempted.

Looking to the future, much work still needs to be completed to
further refine our theoretical understanding. At fixed order, studies
of event shapes at NNLO are just beyond the horizon~\cite{glogar}; in
due course, they will be supplemented by NNL resummation of Sudakov
logarithms~\cite{vogt}. The shape function viewpoint could be
extended in several directions: to more general event shapes in
annihilation processes, which require special treatment within the
dispersive approach, such as jet broadening~\cite{broad},
energy--energy correlations~\cite{eec} and the D
parameter~\cite{dpar}; to other processes, such as DIS, where much is
known already about resummed event shapes, though it was not covered
here because of lack of space~\cite{dasgu}; ultimately, to the more
general and interesting case of QCD hard scattering. Dressed gluon
exponentiation cannot presumably be extended beyond NLL accuracy
without tackling the issue of double renormalon chains, in itself a
question of some theoretical interest. In summary, many interesting
questions remain open, and our understanding of the interface between
perturbative and nonperturbative QCD is likely to become deeper in the
coming years.

As for experiment, this being a brief review form a theoretical
viewpoint, I did not discuss data (see, for example,
Ref.~\cite{exp}). It should however be emphasized that one of the main
reasons for theoretical interest in this area is the remarkable wealth
and precision of the data available, over a wide kinematic range and
for different processes. The precision of the data is such that they
may well be able to discriminate between different current theoretical
models of power corrections in the peak region.  As the work of LEP
collaborations appears to be winding down, it is perhaps worth stating
once more the obvious: it is very important that data and the tools
for their analysis should be preserved in usable form long beyond the
lifetime of the experiments. New theoretical tools that warrant
further data analysis may yet be developed, and good experimental work
done in the past should not be lost.
 
\vspace{1cm}

\noindent {\large {\bf  Acknowledgements}}

\noindent It is a pleasure to thank E. Gardi and G. Korchemsky for 
helpful comments.

\end{document}